\documentclass[12pt,twoside]{article}
\title{Bifurcation Analysis of Liquid Crystal Phase Transitions}
\author{B. M. Mulder \\
                         Institute for Materials Science
                         \thanks{current address: FOM Institute AMOLF,
                                                   P.O. Box 41883,
                                                   1009 DB Amsterdam,
                                                   THE NETHERLANDS} \\
                        N.C.S.R ``Demokritos'' \\
                        15310 Aghia Pareskevi \\
                        GREECE}
\date{Lecture notes, NATO Advanced Research Workshop \\
          Computer Simulation of Liquid Crystals \\
            Il Ciocco,15-21 sept '91}
\usepackage{graphicx}
\newcommand{\be}{\begin{equation}}
\newcommand{\ee}{\end{equation}}
\newcommand{\df}[1]{\mbox{$\rho^{(1)}(#1)$}}
\newcommand{\dfb}{\rho^{(1)}}
\newcommand{\ep}[1]{\mbox{$\epsilon^{#1}$}}
\newcommand{\sm}[1]{{\displaystyle \sum_{#1}^{d-1}}}
\newcommand{\rf}[1]{\mbox{eqn.\ (\ref{#1})}}
\newcommand{\spc}{\hspace{1cm}}
\newcommand{\dd}[3]{\mbox{$\Delta^{(#1)}_{#2,#3}$}}
\begin{document}

\maketitle

\begin{abstract}
These lectures focus on bifurcation analysis as a tool for studying phase
transitions that occur in models of liquid-crystalline systems. We show how
this approach bridges the gap between the phenomenological Landau theory and
the --- often intractable --- full statistical mechanical treatment. Employing
a ``toy model'' as a tutorial example the various ingredients of the technique
are presented. Special attention is paid to the way in which one obtains
information on the relation between the characteristics of the assumed
interparticle interactions (shape, symmetry ...) and global properties of the
phase transitions (order, symmetry of resultant phases ...). Finally a few
more involved examples are discussed indicating how the approach can be applied
to more realistic models and how it can serve as a complement to simulations.
\end{abstract}

\section{Introduction}

Liquid crystalline phase transitions (like most phase transitions) involve the
phenomenon of symmetry breaking. In fact part of their interest derives from
the diversity, and often subtlety, of the ways in which these systems upon
cooling and/or compression stepwise lower their symmetry in order to span the
gap between the high symmetry of the isotropic phase and the ultimate lowest
symmetry crystalline phase. One almost feels that there is an underlying
minimum principle at work that drives these systems to give up as little of
their symmetry as the external conditions allow. Fortunately we posses, in the
form of the Landau theory \footnote{For an up-to-date introduction see
\cite{tole:land} }, a complete descriptive apparatus for symmetry-breaking
phase transitions. The recipe to be followed is quite simple (i) select an
order parameter being an observable that has specified transformation rules
under the symmetry group of the high symmetry phase and whose values
distinguish between the two phases (ii) generate an expansion of the relevant
coarse-grained thermodynamic potential in terms of the order parameter around
the high-symmetry phase. As an example consider the Isotropic-Nematic
transition for which the order parameter is a symmetric, traceless,
three-tensor ${\bf Q}$
\cite{dege:liqc} and the Landau expansion of the free energy takes the form

\be
F = F_{0}+ A Tr{\bf Q}^{2} + B Tr{\bf Q}^3 + C_{1}(Tr{\bf Q}^{2})^{2} +
    C_{2} Tr{\bf Q}^{4} + ...
\ee
Under the assumption that the quadratic coefficient $A$ changes sign at the
transition and the coefficients $B$ and $C_{i}$ are slowly varying we can then
then easily deduce the known properties of the I-N transition by calculating
the minimum the free energy with respect to the order parameter. It should be
noted that this procedure, although powerful, is both phenomenological and
essentially {\em a posteriori}. First of all the selection of the order
parameter either requires prior knowledge of the macroscopic behaviour of the
system or, barring that, a strong dose of physical intuition. One level down
the coefficients in the expansion are either chosen to reproduce the expected
behaviour or simply varied in order to probe the various possible transitions. 

From the molecular point of view, where our only inputs are the particles that
make up our system and the interactions between them, a few obvious questions
immediately arise: (i) can we predict the order parameter and thus the
symmetries of the resultant phase? (ii) can we calculate the analoga of the
expansion coefficients $A, B ...$ that determine the location and the nature of
the transitions? In principle statistical mechanics should supply us with the
answers to these questions. In practice, however, physics (like life itself) is
not as easy as one would wish it to be.  Consider the formulation of
statistical mechanics closest in spirit to the situation at hand: classical
density functional theory (CDFT). In this case we assume that we are supplied
with the relevant thermodynamic potential as a functional of the one-particle
distribution function \df{i}, where $i$ is a shorthand for the degrees of
freedom of a single particle e.g $i = ({\bf r},\Omega)$ for a rigid
non-spherical particle, ${\bf r}$ being the location of its center of mass and
$\Omega$ its orientation with respect to a fixed reference frame. Some crude
analogies with the Landau approach are apparent.  First of all the equilibrium
phase is selected through a variational principle.  The \df{i} codes for the
symmetries of the phase and thus plays the role of the order parameter. Finally
the details of the functional implicitly specify the sought after expansion
coefficients. The first problem we have to confront in this approach is that we
do not know the true functional except possibly in terms of formal expansions,
so we will be forced to make careful approximations that hopefully leave as
much of the relevant physics as possibly. Next, even after making the necessary
approximations, the variational principle will in general yield non-linear
functional relations which are difficult to solve.

Part of the success of the Landau theory is based on the fact that it
exclusively focuses on the description of the system  near phase transitions,
thus singling out the most interesting behaviour at the expense of a more
microscopic description. The density functional formalism on the other hand
does take into account the microscopic degrees of freedom, but in doing so
introduces a level of complexity far beyond that of the Landau theory. The aim
of these lectures will be to show how bifurcation analysis can help to bridge
this gap, by studying the solutions of the variational principle for the
density functional near its critical points. This technique allows us  to
extract the information relevant to two questions posed above i.e to determine
the symmetry of the resultant phases and the nature and location of the phase
transitions. Moreover it often yields to analytical treatment even in cases
where only a minimum amount of information is supplied about the interparticle
interactions e.g just their symmetries, thus giving rise to predictions of a
rather general nature valid for whole classes of systems.

I believe the ideas presented here are useful not just for theorists but also
for those involved in simulating model liquid crystals. There are two areas
where bifurcation analysis can supply information of direct interest to a
simulation:
\begin{description}
\item[description] Solution of even the simplest model having the required
symmetries will often yield the relevant order parameters that can be used
to describe phases that are observed in the simulations.
\item[prediction] Analyzing a simple model for class of related particles
and/or interactions can lead to predictions about the regions in the phase
diagram where interesting behaviour can be expected, thus guiding the choice of
systems to be simulated.
\end{description}

\section{Generalities}

\subsection{Density functional theory}

Starting point of the density functional theory for classical many particle
systems \cite{evan:liqv} is the observation that there exists a functional
${\cal W}[\rho^{(1)}]$ of the one particle distribution function \df{i} with
the following properties:
\begin{enumerate}
\item ${\cal W}[\rho^{(1)}] \geq {\cal W}[\rho^{(1)}_{eq}]$ where
$\rho^{(1)}_{eq}$ is the equilibrium distribution.
\item ${\cal W}[\rho^{(1)}_{eq}] = {\cal W}_{eq}$ where ${\cal W}_{eq}$ is the
thermodynamic equilibrium value of the grand canonical potential.
\end{enumerate}

These two properties together imply a variational principle for obtaining the
full equilibrium properties of the system in question. The big surprise is that
the variation is with respect to a quantity that depends only on the degrees of
freedom of a single particle and not as one would expect with respect to an
N-particle quantity. I should stress  that there is no approximation
involved here, and all many-particle correlations are correctly accounted for.
In fact in a moment I will show how all higher order correlation functions can
be generated from the functional itself. Since there is no such thing as a free
lunch however, we must now face the downside of the theory: We have been told
that the functional exists but have not been given any clue as how to construct
it. Nevertheless we do know that it is has the following general structure
(recall ${\cal W} = F - \mu N$)

\be
{\cal W}[\rho^{(1)}] = {\cal F}[\rho^{(1)}] - \mu \int di \df{i}
\ee
where the second term involving the chemical potential $\mu$ is easily
understood if one remembers that the equilibrium one particle distribution has
the following normalization

\be
\int di \rho^{(1)}_{eq}(i) = N
\label{eq:norm}
\ee
where $N$ is the number of particles in the system. The first term 
represents the free energy of the system and can be expressed as

\be
\beta {\cal F}[\rho^{(1)}] = \int di \df{i} \{ \ln {\cal V}_{T}\df{i} - 1\} - 
            \Phi[\rho^{(1)}]
\ee 
where the first part is the free energy functional for a non-interacting
system, which contains the temperature dependent quantity ${\cal V}_{T}$ being
the thermal volume of the system i.e. the product of the various thermal
wavelengths associated with the kinetic degrees of freedom. As usual the sting
is in the tail, here in the form of the functional $\Phi$ that encapsulates all
contributions due to interactions between the particles. When the particles in
our system interact only pairwise through a potential $v(i,j)$ this functional
can, at least formally, be expanded in a generalized virial series using the
language of diagrams
\cite{hans:simp}

\be
\Phi[\rho^{(1)}] = \left\{ \parbox{50mm}
           {Sum of all connected, irreducible diagrams with $\rho^{(1)}$
            vertices and Mayer function bonds: 
            $f(i,j) = e^{-\beta v(i,j)}-1$} \right.
\label{eq:diag}
\ee
Finally, as promised earlier, this functional can be used to generate all
direct correlation functions through the relation

\be 
c^{(n)}(i_{1},i_{2},...,i_{n};\dfb) = \frac{\delta^{n} \Phi[\rho^{(1)}]}
             {\delta \df{i_{1}} \delta \df{i_{2}} \cdots \delta \df{i_{n}}}
\ee
These in turn can be related through generalized Ornstein-Zernike relations to
the more usual n-particle densities $\rho^{(n)}$ \cite{evan:liqv}, showing that
the density functional  indeed gives full description of N-body equilibrium
system. 

Suppose now that one way or the other (approximations, intuition, reading of
sacred texts \ldots) we have managed to construct a functional relevant
to the system we want to study. As a first step towards solving the variational
problem to determine the equilibrium distribution we then look for the
stationary distributions for which

\be
\frac{\delta {\cal W}[\rho^{(1)}]}{\delta \df{i}} = 0
\ee
Inserting the general form of ${\cal W}[\rho^{(1)}]$ we arrive at the following
selfconsistency relation for the one particle distribution

\be
\df{i} = \frac{1}{{\cal V}_{T}} e^{\beta \mu} 
                    \exp \frac{\delta\Phi[\rho^{(1)}]}
                                            {\delta \rho^{(1)}(i)}
\label{eq:stat}
\ee
revealing the role that the first functional derivative of $\Phi$ plays as a
selfconsistent effective one particle potential. As mentioned in the
introduction this selfconsistency relation that determines the stationary
distributions is a highly non-linear functional relation whose solution even in
the simplest cases requires numerical treatment. The chemical potential $\mu$
is easily eliminated from this equation using the normalization condition 
\rf{eq:norm}.

\subsection{Bifurcation analysis}

As stated before we are not going to try to solve the stationarity equation
\rf{eq:stat} in its full glory but instead concentrate on its behaviour in the
neighbourhood of a symmetry breaking phase transition. The reason that we can
do so is due to the fact that this type of phase transitions is associated
with the appearance of multiple solutions to \rf{eq:stat}. The general
mechanism by which such new solutions appear as the value of some external
parameter in the functional is changed is that of a bifurcation i.e the new
solution branches off from the originally stable solution. The new solution
might immediately be the globally stable one in which case we have a continuous
transition at the bifurcation point or it might be (initially) metastable with
respect to the parent phase in which case one expects a first order transition,
the bifurcation point marking the upper (or lower) limit of stability of the
parent phase.  In Figure~1 we have sketched these two scenarios.
\begin{figure}
\includegraphics[width=\textwidth]{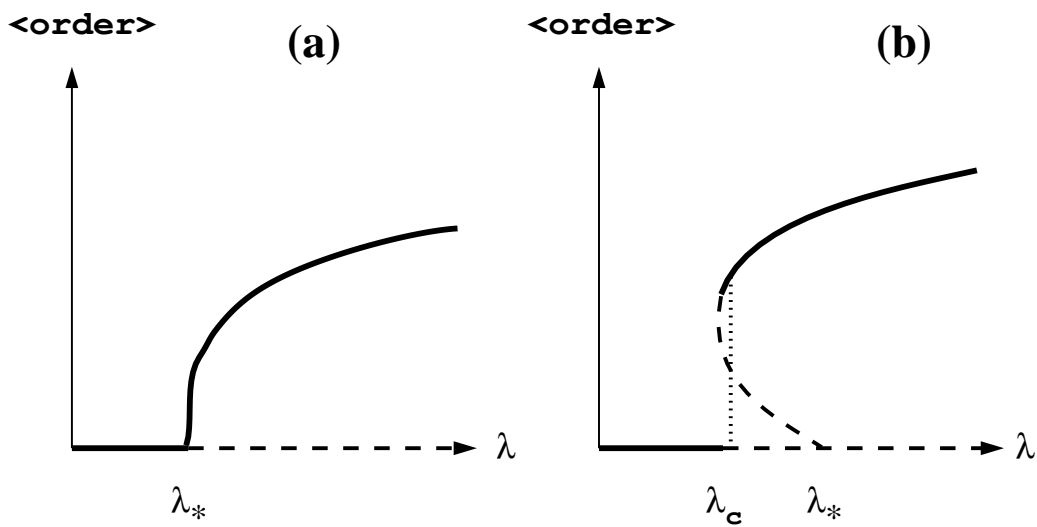}
\caption{Generic bifurcation diagrams. The $x$-coordinate $\lambda$
is the thermodynamic parameter that drives the transition and the
$y$-coordinate a generic order parameter. The thick lines are the stable
solutions while the dashed lines are meta- or unstable solutions. $\lambda_{*}$
identifies the bifurcation point. (a) A continuous transition. The transition
takes places at the bifurcation point (b) A first-order transition. The
transition takes place at $\lambda^{c} \neq \lambda_{*}$.}
\end{figure}
Bifurcation analysis has been developed by mathematicians in order to deal with
these phenomena which are common to many types of non-linear equations and has
found wide application mainly in the field of non-linear differential equations
\cite{chow:bifu}.  Applications to the physics of phase transitions have been
attempted only on a much more modest scale \footnote{for a review see:
\cite{koza:bifu}}, possibly because good examples were much less obvious. By
the end of these lectures, however, I hope to have convinced the reader that
liquid crystalline phase transitions are a ``bifurcators heaven''.

Let us now see how the analysis works in practice. First of all we choose an
external parameter in our functional the variation of which will drive the
system to undergo a phase transition. For definiteness, and since our examples
are all geared towards hard particle systems, we'll take the number density $n
=\frac{N}{V}$ where $V$ is the volume of our system ({\em mutatis mutandi\/}
any other parameter like the temperature, pressure, external fields etc.\ could
have been chosen). Next we need an initially stable solution of the
stationarity equation \rf{eq:stat} that is at least defined for a range of
densities that span the transition, which we'll call the reference solution
$\rho^{(1)}_{0}$. We then look for a solution close the reference solution by
introducing the following expansions in the arbitrary  parameter $\epsilon$
around the, as yet undetermined, density $n_{0}$

\be
n  =  n_{0}+\epsilon n_{1}+\ep{2} n_{2} + ... 
\label{eq:exp1}
\ee
and
\be
\dfb  =  \dfb_{0}+
         \epsilon \{ \dfb_{1}+n_{1} \frac{d\dfb_{0}}{dn}\mid_{n_{0}} \} +
         \ep{2} \{ \dfb_{2} + \frac{1}{2} n_{1}^{2} \frac{d^{2} \dfb_{0}}
         {dn^{2}} \mid_{n_{0}}+ n_{2} \frac{d\dfb_{0}}{dn}\mid_{n_{0}} \} +
         ...
\label{eq:exp2}
\ee 

By separating out the flow along the reference solution, the functions
$\dfb_{l}$ with $l=1,2,...$ represent the actual deviations from the reference
solution at every order in $\epsilon$. Inserting these expansions into the
stationarity equation \rf{eq:stat} and solving order by order in $\epsilon$ we
construct a solution that ``creeps'' along the bifurcating solution branch. The
bifurcation density $n_{0}$ itself will be determined by the lowest order
equation which describes the conditions for the appearance of a non-zero
initial perturbation $\dfb_{1}$. This lowest order equation commonly referred
to as {\em the\/} bifurcation equation turns out to be \footnote{Actually we
have discarded a term here which enforces the correct normalization to zero of
the perturbation for the system in a finite volume. This term, however, vanishes
in the thermodynamic limit where the normalizations are automatically enforced
by symmetry alone}

\be
\label{eq:bif1}
\dfb_{1}(i) = \dfb_{0}(i) \int \!dj\, c^{(2)}(i,j;\dfb_{0})\, \dfb_{1}(j)
\ee
Its structure becomes even clearer if we make the following substitutions
\begin{eqnarray}
\phi_{1}(i) & = & \dfb_{0}(i)^{-\frac{1}{2}}\dfb_{1}(i) \\
K_{0}(i,j)  & = & \dfb_{0}(i)^{\frac{1}{2}}
                         c^{(2)}(i,j;\dfb_{0})\dfb_{0}(j)^{\frac{1}{2}}
\end{eqnarray}
yielding the symmetric representation
\be
\phi_{1}(i) = \int \! dj K_{0}(i,j) \phi_{1}(j)
\label{eq:bifs}
\ee
This is nothing but a generalized eigenfunction equation. Since the kernel
$K_{0}$ depends solely on the properties of the reference phase $\dfb_{0}$ we
have all the necessary information to solve it. The eigenfunctions follow
immediately from the global symmetries of the reference solution since the
kernel $K_{0}$ is invariant under these, which fixes its eigenfunctions. The
bifurcation density is determined as the minimum value of the density for which
$K_{0}$ has an eigenvalue of unity. In the general case there will be a
degenerate set of eigenfunctions satisfying \rf{eq:bifs} so that we will need
more information to fix the true bifurcating eigenfunction(s) in order to
determine the nature of the emerging phase. This information is, as we will see
below, supplied by the next order equation in the hierarchy of bifurcation
equations generated by the expansions \rf{eq:exp1} and \rf{eq:exp2}.

As the higher order equations from the bifurcation hierarchy in the general
case rapidly become rather unwieldy if no appropriate shorthand is introduced,
I will refrain from displaying any but rather make the following comments.
First of all determining the equations is largely a matter of a lot algebra
which can easily be automated using some form of symbolic processing. Second,
and more important, they form a consistent scheme for successively solving for
the unknown perturbations $n_{i}$ and $\dfb_{i}$ i.e. the $k^{th}$-order
equation contains only perturbations of order $\leq k$. Finally, as already
mentioned above, most of the ``juicy'' information already follows from the first
two equations in the hierarchy. That the technique can however be used to
construct the bifurcating solution even quite far away from the bifurcation
point is illustrated in the beautiful and pioneering paper by Kayser and
Ravech{\'{e}} on the Onsager model\cite{kays:onsa}, a paper which in fact
initiated my own involvement with bifurcation analysis.

\subsection{The Onsager approximation}
\label{sec:ons}

To escape slightly from the very general setting described in previous two
subsections we will look somewhat closer at the Onsager approximation for hard
particle systems. This approximation is of course widely used and is to hard
particle systems what mean-field theory is to systems with soft potentials, an
analogy which in fact goes quite deep \cite{kays:onsa,kirk:rect,muld:smec}.
Moreover, as far as universal features of the phase transitions in such systems
is concerned, it seems to be equivalent to a whole class of density functional
theories currently in vogue \cite{muld:bicr}. Dispensing for a moment with its
justification, I just give its formulation in terms of the diagram expansion
\rf{eq:diag}. It consist of keeping just the lowest order term in the
expansion yielding the following form for the free energy density functional

\be
\beta{\cal F}_{Onsager}[\dfb] = \int di \df{i} \{ \ln {\cal V}_{T}\df{i} - 1\}
               + \frac{1}{2} \int \! di \int \! dj \df{i} \df{j} \chi(i,j)
\ee
where I have introduced the characteristic function $\chi(i,j)$ taking on the
value $1$ when the particles overlap and $0$ when they don't, properties
easily deduced from the form of the Mayer function and the fact that we are
dealing with hard potentials. Anticipating some of the examples that are to
follow we will consider only spatially homogeneous phases and restrict our
attention to the orientational degrees of freedom. In this case the singlet
distribution function takes on the form $\df{i} = n \psi(\Omega)$ where $n$ is
the number density and $\psi(\Omega)$ the orientational distribution function
which has unit norm. All integrals over the spatial degrees of freedom can now
be performed turning the Onsager functional into
\begin{eqnarray}
\frac{\beta{\cal F}_{Onsager}[\psi]}{N} &=& 
       \int \! d\Omega \psi(\Omega) \{ \ln \psi(\Omega) -1 \} + \nonumber \\
    & &   \frac{1}{2}n \int \! d\Omega \int \! d\Omega'
       \psi(\Omega)\psi(\Omega') {\cal E}(\Omega,\Omega')+ 
        \ln n{\cal V}_{T}
\end{eqnarray}
which by dividing out the number of particles $N$ allows us to pass painlessly
to the thermodynamic limit. A central role is played by the excluded volume at
fixed orientations

\be
{\cal E}(\Omega,\Omega') = \int \! d{\bf r} \int \! d{\bf r}'
                                \chi({\bf r}'-{\bf r},\Omega,\Omega')
\ee
This role becomes becomes even more clear if we work out the bifurcation
equations in this case for phases developing from the low density isotropic
phase $\psi_{0}(\Omega) = \frac{1}{8\pi^{2}}$. The first two equations are
\be
\label{eq:ob1}
\psi_{1}(\Omega) = -\frac{n_{0}}{8\pi^{2}} \int \! d\Omega'
                   {\cal E}(\Omega,\Omega') \psi_{1}(\Omega')
\ee
and
\begin{eqnarray}
\label{eq:ob2}
\psi_{2}(\Omega) = -\frac{1}{8\pi^{2}}
                  \{ n_{0} \int \! d\Omega' {\cal E}(\Omega,
                  \Omega') \psi_{2}(\Omega')+
                   n_{1} \int \! d\Omega' {\cal E}(\Omega,
                  \Omega') \psi_{1}(\Omega')- \nonumber \\
                  \frac{1}{2} n_{0}^{2}
                  ( \{ \int \! d\Omega' {\cal E}(\Omega,
                  \Omega') \psi_{1}(\Omega') \}^{2}-
                  \frac{1}{8\pi^{2}} \int \! d\Omega' 
                  \{ \int \! d\Omega'' {\cal E}(\Omega',\Omega'')
                  \psi_{1}(\Omega'') \}^{2} ) \}
\end{eqnarray}
showing how the problem, apart from the algebra, reduces to the knowledge of
the eigenvalues and eigenfunctions of the pair excluded volume ${\cal
E}(\Omega,\Omega')$. 

\section{A tutorial example}

\subsection{Rods with restricted orientations}

Rather than continuing the analysis in an abstract setting I would like to work
through a simple example indicating along the way how the results tie in the
general statements made above. Of course by using the big guns on such a
small target one runs the risk of practicing overkill, but I feel the insight
in the method gained through this procedure outweighs this risk. 

The model I consider is the following ``travesty'' of a hard particle fluid:
uniaxial inversion symmetric convex bodies whose symmetry axis can point in a
restricted number of directions namely parallel to the axes of a
$d$-dimensional Cartesian reference system. Allowing general values for the
dimensionality $d$ (rather than just the conventional $d=3$) gives us a bit
more room to play with the model. Such restricted orientation fluids have
already received quite a lot of attention in the past
\cite{zwan:rods,runn:natu,barb:yexp}, so none of the results are going to come
as a surprise. Furthermore, in the light of the remarks made in section
\ref{sec:ons}, the discussion will be restricted to the Onsager approximation.

Let us label the allowed orientations of the particles by $s$, where it will
turn out to be convenient to let $s$ range from $0$ to $d-1$. The excluded
volume between two particles with fixed orientations has an exceedingly simple
form 

\be
{\cal E}(s,s') = e_{\parallel}\delta(s,s')+e_{\perp}(1-\delta(s,s'))
\ee
Because the particles are convex and non-spherical the excluded volume
,$e_{\parallel}$, when they are parallel is smaller than the corresponding
quantity, $e_{\perp}$, when they are not. Remember that all the
orientations are mutually orthogonal so that, given the symmetry of the
particles, all non-parallel directions are equivalent as far as the excluded
volume is concerned. Introducing the dimensionless number density $\eta = n
(e_{\perp}-e_{\parallel})$ we can write down the density functional as

\begin{eqnarray}
\label{eq:ex0}
\frac{\beta{\cal W}[\psi]}{N} &=& \sm{s=0} \psi(s)\{\ln\psi(s)-1\}-
\frac{1}{2}\eta \sm{s=0} \sm{s'=0} \delta(s,s')\psi(s)\psi(s')+
             \nonumber \\
 & & \ln {\cal V}_{T}n +\frac{1}{2}n e_{\perp}-\beta\mu \sm{s=0} \psi(s)
\end{eqnarray}
The chemical potential here just serves as a Lagrange multiplier which is used
to obtain the correct normalization of the orientational distribution, and will
be eliminated immediately. We therefore find, after performing the variation
with respect to $\psi$, the following selfconsistency equation

\be
\psi(s) = \frac{\exp{\eta \sm{s'=0} \delta(s,s')\psi(s')}}
               {\sm{s'=0} \exp{\eta \sm{s''=0} \delta(s',s'') \psi(s'')}}
\ee
This equation might be {\em deja vu \/} for some readers, since it is nothing
but the mean-field equation for the d-state Potts model on an arbitrary lattice
if we identify $\eta = \beta zJ$ where $J$ is the coupling constant and $z$ the
coordination number of the lattice. Note that the isotropic solution $\psi_{0}
= \frac{1}{d} $ which plays the role of reference phase, is a solution at all
densities.

\subsection{Analysis}
Instead of rederiving the bifurcation equations from the start we can use the
general results of section \ref{sec:ons} for the Onsager case, if we make the
following changes: (i) replace all integrations over the orientation $\Omega$
by sums over the discrete variables $s$ (ii) replace every factor $8\pi^{2}$ by
$d$ being the ``volume'' of the discrete orientation space and (iii) replace
the excluded volume ${\cal E}(\Omega,\Omega')$ by $-\delta(s,s')$ and, finally,
(iv) change all references to the number density $n$ into the dimensionless
density $\eta$.  This results in

\be
\label{eq:ex1}
\psi_{1}(s) = \frac{\eta_{0}}{d} \sm{s'=0}
                   \delta(s,s') \psi_{1}(s')
\ee
and
\begin{eqnarray}
\label{eq:ex2}
\psi_{2}(s) &=& \frac{1}{d}
                  \{ \eta_{0} \sm{s'=0} \delta(s,s') \psi_{2}(s')+
                     \eta_{1} \sm{s'=0} \delta(s,s') \psi_{1}(s')+ \nonumber \\
            & &     \frac{1}{2} \eta_{0}^{2}
                  ( \{ \sm{s'=0} \delta(s,s') \psi_{1}(s') \}^{2}-
                    \frac{1}{d} \sm{s'=0} \{ \sm{s''=0} \delta(s',s'')
                                        \psi_{1}(s'') \}^{2} ) \}
\end{eqnarray}
which are the analogs of the equations \rf{eq:ob1} and \rf{eq:ob2}.

Ignoring for the moment that first bifurcation equation is actually trivial in
this case (we are faced with the daunting task of diagonalizing the identity
matrix !), we are going to take a round-about way by exploiting the symmetries
of the reduced excluded volume $\delta(s,s')$ to obtain a complete set of
eigenfunctions. This procedure prepares the way for later applications where
the symmetries of the excluded volume will play a crucial role in the analysis.
Instead of exploiting the full symmetry, which is that of the symmetric group
$S_{d}$ of all permutations of $d$ objects, we can get away with just the
subgroup formed by the cyclic permutations $C_{d}$. The irreducible
representations of this abelian group are the functions

\be
\phi_{k}(s) = e^{\frac{2\pi i}{d} k s} \spc  k = 0,1,\ldots,d-1
\ee
They form an orthogonal set under the innerproduct defined by
\be
\langle \phi_{i},\phi_{j}\rangle = \sm{s=0}\phi_{i}^{*}(s)\phi_{j}(s)
                                   = d \,\delta(i,j) 
\ee
where the $*$ denotes complex conjugation. The reduced excluded volume has the
simple expansion

\be
\delta(s,s') = \frac{1}{d} \sm{k=0}\phi_{k}(s)\phi_{k}^{*}(s')
\ee
which is the completeness relation for the irreps of $C_{d}$. Combining these
two relations one sees that every $\phi_{k}$ with $k=0,1,\ldots,d-1$ is an
eigenfunction of the reduced excluded volume with eigenvalue unity. Note,
however, that $\phi_{0}$, being the identity representation of $C_{d}$, is just
a constant and therefore proportional to the isotropic distribution $\psi_{0}$,
so should not be included in the bifurcating eigenfunction, whose general form
thus becomes
\be
\psi_{1}(s) = \sm{k=1}c_{k}\phi_{k}(s)
\ee
Inserting this form into the bifurcation equation \rf{eq:ex1} one immediately
obtains the bifurcation condition
\be
\frac{\eta_{0}}{d} = 1
\ee
which fixes the bifurcation density $\eta_{0} = d$. We now know {\em when\/}
the bifurcation occurs, but are still in the dark as to {\em what \/} exactly
happens, since the coefficients $c_{k}$ in the general form of the bifurcating
eigenfunction are as yet undetermined, reflecting the fact that the $\phi_{k}$
form a degenerate set of eigenfunctions of the reduced excluded volume.  Having
exhausted the first bifurcation equation, this is clearly the point where the
second bifurcation equation \rf{eq:ex2} comes in. The fact that this equation
can be used for the purpose at hand {\em without solving for its unknowns} is
due to two surprises. The first surprise is that we can eliminate the unknown
second order perturbation $\psi_{2}$. The key ingredient to this elimination is
the following identity
\be
\langle \phi_{k},\psi_{2}\rangle=\frac{\eta_{0}}{d}\sm{s=0}\sm{s'=0}
               \phi_{k}^{*}(s)\delta(s,s')\psi_{2}(s')
\ee
which is valid for any $k=0,1,\ldots,d-1$. This identity, which one easily
checks in this special case, follows from a general property of the excluded
volume, namely that it is invariant under the interchange of the two particles
involved (in more formal terms this means that it can be interpreted as a
hermitian operator on the space of single particle distributions equipped with
a suitable innerproduct). The recipe is now as follows: take the innerproduct
of \rf{eq:ex2} with any of the $\phi_{k}$ with $k=1,\ldots,d-1$ and use the
identity given above to equate the left hand side to the first term on the
right hand side. This leaves $d-1$ equations involving the unknowns $c_{k}$ and
$\eta_{1}$. Skipping the intermediate algebra these equations can be written as
\be
\label{eq:ex4}
\frac{2\eta_{1}}{d^{2}}c_{k}=
                 -\sm{l=1}\sm{m=1}\delta_{\mbox{$k,(l+m) \bmod d$}}\:c_{l}c_{m}
\ee
The second surprise is that one can scale away the prefactor containing the
unknown first perturbation in the density $\eta_{1}$ (provided it is non-zero,
of course) by changing to variables $b_{k} = - \frac{d^{2}}{2\eta_{1}}c_{k}$
yielding the simplified equation
\be
b_{k} = \sm{l=1}\sm{m=1}\delta_{\mbox{$k,(l+m) \bmod d$}}\:b_{l}b_{m}
\ee
This equation is not as bad as they come and  we can find $d$ solutions to it
labeled by $n=0,1,\ldots,d-1$
\be
b_{k}^{(n)} = \frac{1}{d-2}e^{\frac{2\pi i}{d} n k} 
\ee
Putting it all together we find $d$ acceptable bifurcating eigenfunctions
$\psi_{1}$ 
\begin{eqnarray}
\psi_{1}^{(n)}(s) &=& \sm{k=1}c_{k}^{(n)}\phi_{k}(s) \nonumber \\
                  &=& -\frac{2\eta_{1}}{d^{2}(d-2)}\sm{k=1}
                                e^{\frac{2\pi}{d} i k (n-s)} \nonumber \\
                  &=& -\frac{2\eta_{1}}{d^{2}(d-2)}(d\delta(n,s)-1) 
\end{eqnarray}
The last identity brings us to our goal since it shows that the bifurcating
solution is uniaxially symmetric about the ordering axis labelled by $n$, the
fact that there are $d$ such solutions simply reflects the fact that the axes
are all equivalent. We can thus conclude that we are dealing with an isotropic
to nematic transition. Since we have assumed that $\eta_{1} \neq 0$, we must be
dealing with a first order transition (the case depicted in fig.1.b). Indeed
in order for $\psi_{1}$ to represent enhancement of order in a certain
direction we must have $\eta_{1} < 0$ consistent with the ``bending back'' of
the solution i.e. the creation of a v.d. Waals loop in the equation of state.
The solution presented here is clearly valid only for $d \geq 3$. The case
$d=2$ is special and a glance at the equations shows that in this case we must
have $\eta_{1}=0$, the fingerprint of a continuous transition (see fig 1.a).
This should come as no surprise since the model for $d=2$ is nothing but the
Ising model, in one of its many disguises.

This is a good point to reflect on what we have achieved so far. Starting from
the defining equations of our model we have derived by purely analytical means
the location of the bifurcation point, an upper limit to the stability of the
isotropic phase, as a function of the dimension and the parameters
$e_{\parallel}$ and $e_{\perp}$ that describe the interactions between the
particles involved. Moreover we have determined the order of the phase
transition involved as well as the nature of the resultant phase. ``Big deal,
most of this was intuitively clear anyway'', I hear the skeptical reader say.
Very true, of course, for the extremely simple model discussed here. However,
the method, although devoid of intuition, is also free of prejudice and works
just as well in more complicated situations where intuition might not be of any
help. I also hope that the reader has gotten some flavor of how the method
focusses on rather general properties of the model being studied; most
conclusions follow from the properties of the reduced excluded volume which is
a quantity heavily constrained by symmetry- and other physical requirements. It
is this feature which allows it to deal with whole classes of particles and/or
interactions many detailed features of which need not be given in order to
obtain the type of results we are after.

\subsection{Connection with Landau theory}
As mentioned in the introduction the combination of density functional theory
and bifurcation analysis shows some analogy to the Landau theory of
phase transitions. I would here like to pursue this analogy in some detail
for the ``toy model'' just introduced. The bridge between the two theories is
formed by the invariant expansion the distribution function involved. In our
case where we have assumed a global $C_{d}$ symmetry the correct set of basis
functions are the irreps $\phi_{k}$ defined in the previous subsection. The
general form of the distribution function therefore is
\be
\psi(s) = \sm{k=0}a_{k}\phi_{k}(s)
\ee
Since $\psi$ must be normalized to unity we have the following constraint on
the expansion coefficients $a_{k}$
\be
\sm{s=0}\psi(s) = \langle \phi_{0},\psi(s) \rangle = da_{0} =1
\ee
This leaves the set $\{a_{k}\}_{k=1,\ldots,d-1}$ as free parameters. One more
constraint is the fact that the distribution function must be real yielding the
relation $a_{k}^{*}=a_{d-k}$. 

The isotropic phase $\psi_{0}$ is characterized by
\be
a_{0} = \frac{1}{d} \;, \spc  a_{k} = 0 \spc  k=1,\ldots,d-1
\ee
making the set $\{a_{k}\}_{k=1,\ldots,d-1}$ a good candidate for an order
parameter.  By construction they also transform irreducibly under the symmetry
group $C_{d}$ of the isotropic so that they indeed form a set of order
parameters in the sense of Landau \cite{tole:land}. The next step is to
introduce the expanded form of the distribution function in the free energy
functional \rf{eq:ex0} and expand with respect to the order parameters
assuming that they are small, as is the case near a phase transition. To third
order in the $a_{k}$ we find
\begin{eqnarray}
\frac{\beta {\cal F}}{N} \equiv f &=&  f_{0} +
    \frac{1}{2}d(d-\eta)\sm{k=1}a_{k}^{*}a_{k}- \nonumber \\
      & & \frac{1}{6}d^2\sm{k=1}\sm{k'=1}\sm{k''=1}a_{k}a_{k'}a_{k''}
    \delta_{(k+k') \bmod d,d-k''}+ \ldots
\end{eqnarray}
From the vanishing of the coefficient of the term quadratic in the order
parameters we immediately recover the bifurcation condition $\eta=d$. Moreover,
if we ignore for the moment the presence and influence of higher order terms in
the expansion and differentiate with respect to the $a_{k}$ in order to
minimize the free energy, we recover equations  equivalent to \rf{eq:ex4},
leading to the result that the solution has the expected uniaxial symmetry. In
this way one obtains exactly the same information as one got from the first two
bifurcation equations. The last step, however, can hardly be called systematic,
and one would really need a more sophisticated analysis in terms of the
algebraically independent invariants along the lines of Prokrovskii and Kats
\cite{prok:land} to establish the claimed result. The source of this problem is
the fact that we are expanding the functional at a fixed value of the density.
The density, in contrast to the $\epsilon$-parameter in the bifurcation
analysis, is not such a good measure of the distance to the bifurcation and
does not allow us to separate the succesive perturbations to the reference
phase that determine the properties of the emergent phase.

Expanding the free energy functional in a suitable set of order parameters thus
leads to a problem {\em formally equivalent\/} to the Landau expansion but with
the {\em big difference\/} that the coefficients in the expansion explicitly
contain microscopic information about the particles and their interactions.
Although the procedure outlined above leads to the same information, I feel
that from a calculational point of view the systematics of the bifurcation
analysis applied to the stationarity equations rather than to the functional
are clearly to be preferred.

\section{Applications}

In this section we will look at some applications of bifurcation analysis to
more (or less) realistic models of liquid crystals. No attempt is made to
review all aspects of the models discussed, but rather to indicate how the
analysis reveals the salient aspects of the phase transitions involved and how
these compare to results obtained by simulations. First of all the ``nematic''
to smectic~A transition in a system of perfectly aligned hard rods is
discussed. Although perhaps a somewhat artificial model it is nevertheless
historically important in the sense that it was the first system for which
conclusive evidence \cite{stro:smec} was obtained that purely repulsive
interactions can lead to liquid crystalline phases beyond the traditional
nematics, a possibility that up till then had been actively dismissed by most
workers in the field. This discovery has given a new lease on life on hard
particle models in liquid crystal research, which is a welcome development both
for simulators and theorists alike. Next the class of biaxial hard particles is
taken on. Here the analysis really comes alive, since it allows us to infer
many important properties of the phase diagram without resorting to a specific
calculation on a single model. Finally I look towards the future and discuss
some current projects and thoughts about future developments.

\subsection{Parallel hard rods}

Consider a fluid of hard cylinders perfectly aligned along a given direction
which we will identify with the $z$-axis of our coordinate system. Without
further justification we adopt the Onsager approximation \cite{muld:smec}. More
elaborate functionals have been constructed for this model
\cite{wenm:smec,poni:nsa}, but these do not lead to qualitatively different
results. The characteristic function of the excluded volume of two cylinders at
a relative separation ${\bf r} = {\bf r}_{2}-{\bf r}_{1}$ is given by
\be
\chi({\bf r}) = \Theta(\sigma^2-x^2-y^2)\Theta(L-|z|)
\ee
where $\sigma$ and $L$ are the diameter and the length of the cylinders
respectively and $\Theta(\cdot)$ denotes the Heavyside function. The analog of
the first bifurcation equation \rf{eq:bif1} is given by
\be
\rho_{1}({\bf r}) = - n \int \!d{\bf r}'\chi({\bf r}-{\bf r}')\rho_{1}({\bf r}')
\ee
where as before $n$ is the number density of the spatially homogeneous
``nematic'' phase formed by the low density system. Since the characteristic
function is invariant under global translations, the sought after
eigenfunction must be a plane wave $\phi_{{\bf q}}({\bf r}) = \exp i
{\bf q \cdot r}$. Inserting this into the equation we find the bifurcation
condition 
\be
1 = - 2\pi\sigma^{2}Ln j_{0}(q_{\parallel}L)
       \left(\frac{J_{1}(q_{\perp}\sigma)}{\frac{1}{2}q_{\perp}\sigma}\right)
\ee
where $j_{0}$ is a spherical Bessel function and $J_{1}$ an ordinary one and we
have decomposed the wavevectors in its components along and perpendicular to
the alignment axis. This equation as it stands is heavily underdetermined.
Fortunately we can use the physical requirement that we should look for the
{\em smallest} density at which there exists a solution, since this is where
the low density spatially disordered phase becomes unstable. Given this
requirement there is just one relevant solution
\begin{eqnarray}
n_{0} &=& 0.7321 \sigma^{-2}L^{-1} \nonumber \\
q_{\parallel,0} &=& 4.493 L^{-1}  \nonumber \\
q_{\perp,0} &=& 0
\end{eqnarray}
which describes the onset of a smectic density wave along the alignment axis.
Knowing the nature of the bifurcating solution simplifies the further analysis
because it allows us to parametrize the one particle density as
\be
\rho({\bf r}) = n \left(1+\sum_{l=1}^{\infty}a_{l}cos(lqz)\right)
\ee
In order to perform the bifurcation analysis to higher order we insert this
parametrization into the stationarity equation and make the following 
expansions
\begin{eqnarray}
n &=& n_{0}+\epsilon n_{2}+\epsilon^{2}n_{2}+\ldots \nonumber \\
a_{l} &=& a_{l,0}+\epsilon a_{l,1}+\epsilon^{2} a_{l,2}+\ldots \spc
l=1,2,\ldots \nonumber \\
q &=& q_{0}+\epsilon q_{1}+\epsilon^{2} q_{2}+\ldots
\end{eqnarray}
Without going into details I just quote the most important results: (i) The
free energy of the smectic phase for $n > n_{0}$ is indeed lower than that of
the nematic phase, showing that the bifurcation leads to a thermodynamically
stable phase (ii) $n_{1} = 0$, so the predicted phase transition is second
order in full agreement with the simulation results. Of course, the Onsager
approximation is too crude to get quantitative results comparable to the
simulations. One improvement suggested in ref. \cite{muld:smec} is to add more
terms in the diagram expansion of the free-energy. Carrying out this program up
to the fourth order diagrams one finds the following values for the critical
packing fraction $\eta = \frac{1}{4}\pi\sigma^{2}L n$ and the wavelength
$\lambda$ of the smectic modulation at the transition
\be
\eta_{c}^{(4)} = 0.37  \spc        \lambda_{c}^{(4)} = 1.34 L
\ee
These values compare favourably to the simulation results
\be
\eta_{c}^{MC} = 0.36  \spc        \lambda_{c}^{MC} = 1.27 L
\ee

\subsection{Biaxial particles}

Most of the convex hard particles which have been studied as models for liquid
crystals have the property of being uniaxially symmetric i.e possess an axis of
rotational symmetry. They therefore seem doomed to form, if anything, nematic
phases which also posess this same symmetry, at least as the first stage of
symmetry breaking from the low-density isotropic phase. Particles in general,
however, are not uniaxially symmetric. How does this influence the formation of
the nematic phase? Under what conditions can phases with lower symmetry develop
from the isotropic phase? It is this type of questions that is ideally suited
for an attack by bifurcation analysis \footnote{In fact there exists a large,
but unfortunately rather impenetrable, mathematical bibliography on the
relationship between symmetry breaking and bifurcations. For a recent review
supposedly aimed at a more physical audience see \cite{gaet:bifu}}. The example
to be discussed here is what happens if the particles have biaxial symmetry of
the type $D_{2h}$, which is the symmetry group of a rectangular box with at
least one of its sides different in length than the other two. Since $D_{2h}$
is a subgroup of $C_{\infty h}$, this class of particles contains all the
uniaxially symmetric, inversion invariant, models as well so comparisons with
previous results are easy. Examples of such particles are general ellipsoids
and sphero-platelets \cite{muld:sphp} which have the well-studied ellipsoids of
revolution and sphero-cylinders as special cases respectively. We again call on
the Onsager approximation to illustrate the working. This is not really a heavy
restriction since I have recently shown \cite{muld:bicr} that the results about
to be presented hold without change for a large class of functionals that
comprises most of the ones proposed in the literature e.g. scaled particle
theory \cite{cott:scal} and more recently the Smoothed Density Approximation
\cite{holy:smoo}. The common element of all these approximations is that the
dependence of the excess free energy of the system on the orientational
distribution of the molecules is described in terms of the distribution
averaged excluded volume
\be
{\cal E}_{ave}[\psi] = \frac{1}{2} \int \! d\Omega_{1} \int \! d\Omega_{2}
       \psi(\Omega_{1})\psi(\Omega_{2}) {\cal E}(\Omega_{1},\Omega_{2})
\ee
which is also the second virial coefficient in a density expansion of the free
energy. As it turns out all symmetry related properties of the phases that
develop from the isotropic phase are determined {\em solely} by the properties
of the excluded volume ${\cal E}(\Omega_{1},\Omega_{2})$ regardless of the
precise form of the functional. Of course the more ``non-universal'' features
of the transitions, like the location of the bifurcations, do depend on the
specific functional.

Let's start then by analyzing the properties of the excluded volume of two
particles with fixed orientations  that follow from symmetry considerations
alone. First of all global rotational invariance dictates that it should be a
single argument function $E$  of the relative orientation $\Omega =
\Omega_{2}^{-1}\Omega_{1}$ of the two particles only, or  
\be
{\cal E}(\Omega_{1},\Omega_{2}) = E(\Omega)
\ee
Symmetry with respect to the interchange of the two particles involved imply
that the function $E(\Omega)$ is invariant under taking the inverse of its
argument 
\be
E(\Omega^{-1}) = E(\Omega)
\ee
This last property, together with the fact that the excluded volume is a real
quantity, implies that the excluded volume interpreted as an operator using the
following prescription
\be
\label{eq:op}
{\cal E}[f](\Omega_{1}) = \int \!d\Omega_{2}\,{\cal E}(\Omega_{1},\Omega_{2})
f(\Omega_{2}) 
\ee
is hermitian on the space of real valued functions of orientation equipped with
the following innerproduct
\be
\langle f,g \rangle = \int \!d\Omega\, f(\Omega)g(\Omega)
\ee
This property, as mentioned in the section on the tutorial example, is a
crucial ingredient of the analysis. Finally we have to implement the $D_{2h}$
symmetry of the particles themselves. If $g_{1}$ and $g_{2}$ arbitrary elements
of this symmetry group interpreted as a rotation and/or inversion, then we must
require the following identity
\be
{\cal E}(\Omega_{1}g_{1},\Omega_{2}g_{2}) = {\cal E}(\Omega_{1},\Omega_{2})
\ee
to hold, or equivalently
\be
E(g_{2}^{-1}\Omega g_{1}) = E(\Omega)
\ee
A set of functions of the relative orientation that have the above symmetries
can be obtained from the usual Wigner rotation matrices $D^{(l)}_{m,n}$ by the
following projection
\be
\dd{l}{m}{n}(\Omega) \propto  \sum_{g_{1} \in D_{2h}} \sum_{g_{2} \in D_{2h}}
               D^{(l)}_{m,n}(g_{2}^{-1}\Omega g_{1})
\ee
which if one works this out yields the, suitably normalized, functions
\begin{eqnarray}
\dd{l}{m}{n} = (\frac{1}{2} \sqrt 2)^{2+\delta_{m,0}+\delta_{n,0}}
               \left(D^{(l)}_{m,n}+D^{(l)}_{-m,n}+D^{(l)}_{m,-n}+
                      D^{(l)}_{-m,-n}\right) \nonumber \\
l = \mbox{even} \spc m,n \geq 0 \mbox{\ and even}
\end{eqnarray}
Putting it all together we can expand the excluded volume as
\be
{\cal E}(\Omega_{1},\Omega_{2}) = {\sum_{l,m,n}}'\: \frac{(2l+1)}{8\pi^{2}}
                         E_{l,mn}\dd{l}{m}{n}(\Omega_{2}^{-1}\Omega_{1})
\ee
where the prime reminds us of the restrictions on the indices $l,m$ and $n$.
Moreover, the particle interchange symmetry implies that the expansion
coefficients are symmetric in the indices $m$ and $n$ i.e $E_{l,mn} =
E_{l,nm}$. In this way we have milked all the information from the various
symmetry constraints that apply to the excluded volume achieving a considerable
reduction in the number of parameters that enter into the problem.

The next step in the program is to solve the lowest order bifurcation equation
(cf. \rf{eq:ob1}). This involves some explicit properties of the functions
$\dd{l}{m}{n}$, the details of which need not concern us here. The most
important point is that these functions for a fixed value of the angular
momentum index $l$ form $(\frac{l}{2}+1)^{2}$-dimensional invariant subspaces
under the operation \rf{eq:op} and that every eigenvalue is
$(\frac{l}{2}+1)$-fold degenerate. For reasons explained earlier on, we are
looking for the eigenvalue that will yield the smallest value of the
bifurcation density. Given the assumptions made --- convexity, and more
importantly pure $D_{2h}$ symmetry so {\em no} cubic symmetry --- this
relevant eigenvalue will be found in the subspace with $l=2$, which describes
the most coarse scale deviations from isotropicity. The result for the
bifurcation density is
\be
n_{0} = -\frac{8\pi^{2}}{\frac{1}{2}(E_{2,00}+E_{2,22})-\frac{1}{2}
         \sqrt{(E_{2,00}+E_{2,22})^{2}+4E_{2,02}^{2}}}
\ee
while the two degenerate eigenvectors $\phi_{0}$ and $\phi_{2}$ are given by
\begin{eqnarray}
\phi_{0} &=& e_{0} \dd{2}{0}{0}+e_{2} \dd{2}{0}{2} \nonumber \\
\phi_{2} &=& e_{0} \dd{2}{2}{0}+e_{2} \dd{2}{2}{2} 
\end{eqnarray}
where the coefficients are the following explicit functions of the expansion
coefficients 
\begin{eqnarray}
e_{0} = -\frac{E_{2,02}}{\sqrt{E_{2,02}^{2}+\tau^{2}}} \spc
e_{2} = \frac{\tau}{\sqrt{E_{2,02}^{2}+\tau^{2}}} \nonumber \\
\tau = \frac{1}{2}(E_{2,00}-E_{2,22})+\frac{1}{2}
         \sqrt{(E_{2,00}-E_{2,22})^{2}+4E_{2,02}^{2}}
\end{eqnarray}
Note that the bifurcation density and the eigenfunctions are completely
determined by the three expansion coefficients $E_{2,00}, E_{2,02}$ and
$E_{2,22}$ of the excluded volume. This is reasonable since, intuitively, a
convex particle of $D_{2h}$ symmetry has three independent dimensions that fix
its coarse-scale shape (cf. the side lengths $a,b$ and $c$ of a rectangular
box). Since the absolute volume of the particle is irrelevant, and can be
absorbed into a redefined density, there are effectively only two free
parameters that describe the specific shape (for the rectangular box one could
take the ratios $\frac{a}{c}$ and $\frac{b}{c}$ for instance).

Finally, we have to determine the actual bifurcating eigenfunction, in order to
learn what the symmetry of the new phase is. We know
that it is a linear combination of the degenerate eigenfunctions obtained from
the lowest order bifurcation equation
\be
\psi_{1} = c_{0} \phi_{0}+ c_{2} \phi_{2}
\ee
Following the procedure already outlined in the tutorial example we can
determine the unknown coefficients $c_{0}$ and $c_{2}$ by using the second
order bifurcation equation \rf{eq:ob2}. The interested reader can find the
details in the original reference \cite{muld:biax}. Amazingly enough, the
result depends only on the sign of a single quantity
\be
\nu = e_{0}(e_{0}^{2}-3e_{2}^{2})
\ee
where the $e_{n}$ are the components of the degenerate
eigenfunctions $\phi_{m}$ on the basis $\dd{2}{m}{n}$. We distinguish the
following cases
\begin{description}
\item[$\nu > 0$] The solution has uniaxial symmetry and describes the ordering
of the major axis of the particle in a preferential direction i.e a rod-like
nematic phase which we denote by $N_{(+)}$. The transition to this phase will
be of first order.
\item[$\nu = 0$] The solution has the $D_{2h}$ symmetry of a biaxial
nematic phase $N_{(biax)}$. The transition to this phase is of second order.
\item[$\nu < 0$] The solution again has uniaxial symmetry but now describes the
ordering of the minor axis of the particle in a preferential direction i.e a
disk-like nematic phase $N_{(-)}$. The transition is again of first order.
\end{description}

The particles for which $\nu = 0$ form lines of Landau bicritical points in the
shape-density phase diagram being the endpoints of the first order transition
lines to the rod- and disk-like nematic phases and marking the four-phase
coexistence of the isotropic, rod-like, disk-like and biaxial phases. 
The solutions to the equation $\nu=0$ can be given explicitly in terms of the
excluded volume expansion coefficients $E_{2,mn}$ as
\be
E_{2,02} = 0, \spc E_{2,00}-E_{2,22} > 0 
\ee
and
\be
|E_{2,02}| = -\frac{1}{2} \sqrt{3}(E_{2,00}-E_{2,22})
\ee
The fact that there are two equations reflects the underlying arbitrariness of
the choice of the remaining two axes of the particle fixed frame once the
primary axis is chosen. A generic impression of a slice of the phase diagram in
the neighbourhood of such a point is sketched in figure 2.
\begin{figure}
\includegraphics[width=\textwidth]{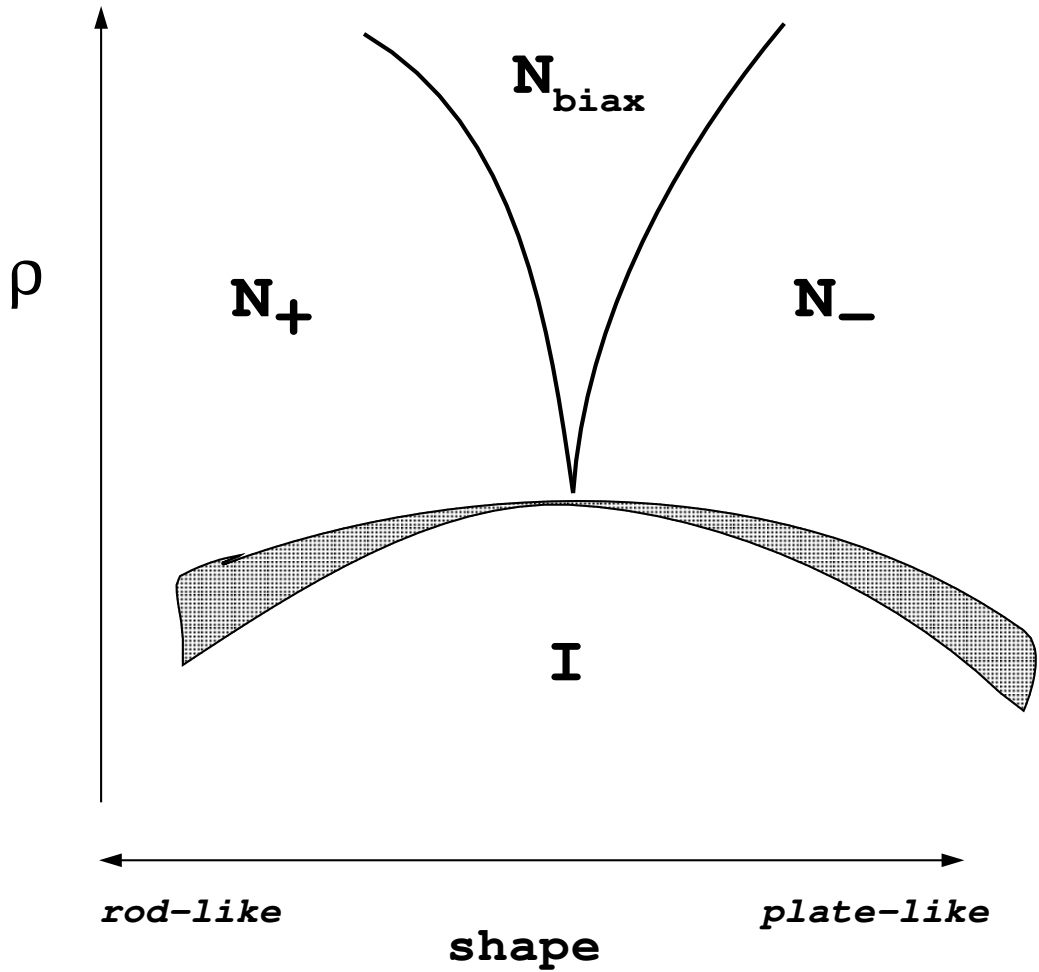}
\caption{Generic phase diagram in the neighbourhood of the
$I-N_{(+)}-N_{(biax)}-N_{(-)}$ multicritical point. x-axis: parameter
describing shape of particle. y-axis: density. The full lines are continuous
transitions while the dashed lines bound the coexistence region of first order
transitions.} 
\end{figure}
The existence of these bicritical points can be understood by considering the
following ``thought experiment'' on a system of rectangular blocks with two
side lengths $c > a$ considered fixed and third $b$ considered variable. We
start with $b=a$ where the particle is an effectively uniaxial rod clearly
disposed to forming a rod-like nematic. If we now gradually increase $b$ until
$b=c$ we end up again with an effectively uniaxial shape but now clearly
disk-like and bound to form a nematic phase where the normal to the disk will
be ordered. The two types of nematic order $N_{(+)}$ and $N_{(-)}$ cannot be
transformed continuously into each other so another phase must intervene, which
perforce has a lower symmetry. This leads to the conclusion that there must be
at least one intermediate value of $b$ for which the particle is neither enough
rod-like nor disk-like to form the corresponding phases. A duality argument
first proposed by Straley \cite{stra:biax} which maps rods into equivalent
disks can then then be used to show that there is a unique value of $b$ for
which this intermediate phase is accessible from the isotropic phase.

Sphero-platelets are to date the only particles for which the expansion
coefficients $E_{2,mn}$ have been calculated analytically and the equation
$\nu=0$ determining the ``bicritical'' particles solved explicitly. This
solution suggests strongly that asymptotically in a regime where the largest
dimension of the particle is much larger than the smallest dimension ($c \ll
a$) the intermediate dimension $b_{*}$ of the ``bicritical'' particle is
approximately the geometric mean of the other two
\be
b_{*} \sim \sqrt{ac}
\ee
This prediction was recently verified in the first extensive simulation on a
system of biaxial particles ---in this case general ellipsoids--- performed by
Allen \cite{alle:biax}. This result is all the more noteworthy since it shows
that the results obtained on these symmetry related questions in the type of
approximations treated here remain relevant to the full statistical mechanics
of the problem. This supports some of the ideas I have presented regarding the
``universality'' of results obtained from the bifurcation analysis even of
highly approximate free energy functionals.

In summary we have managed to determine for a whole class of convex hard
particles the properties of the phases that are reached by phase transitions
from the low density isotropic phase, as predicted by a whole class of
free-energy functionals. In order to apply the analysis to any specific
particle in this class one needs as input only three numbers:
$E_{2,00},E_{2,02}$ and $E_{2,22}$. It is precisely this generality and economy
of description, which focusses only on those parameters in the problem which
are relevant to the properties of the actual transitions, which make
bifurcation analysis such an attractive tool in the study of symmetry-breaking
phase transitions.

\subsection{\ldots and beyond}

The question of which problem to tackle next using the techniques described
here is difficult since we are faced with an {\em embarras de choix}. The
phenomenology of liquid crystals has grown so immensely during the last two
decades that the possibilities seem inexhaustible. I'll therefore restrict
myself to two directions. The first is interesting also from a methodological
point view, while the second concerns qualitatively new and hitherto unexpected
phases.

The first category of problems concerns the transitions from already ordered
phases. All the examples treated in these lectures were transitions from the
totally disordered state. The general theory, however, deals equally well with
these order-order phenomena. Good examples are the nematic-smectic transition
in a system of freely rotating rods and the as yet not completely understood
sequence of phase transitions in the parallel hard cylinder- and spherocylinder
systems. Both these problems have already been studied in the past ($N-S_{A}$
transition: \cite{stec:nsa,poni:nsa}, parallel hard rods:
\cite{tayl:para,holy:para}), but I believe the last word has not been spoken
yet. In the case of the nematic-smectic transition, for instance, it has up to
now always been assumed that the smectic fluctuation that appears at the
transition is decoupled from the orientational order.  Technically speaking
this means that a smectic fluctuation of the form
\be
\psi_{1}(z,{\bf n}) = \psi_{0}({\bf n}) \cos qz
\ee
is introduced {\em by hand}, where ${\bf n}$ denotes the unit vector
along the particle's symmetry axis and $\psi_{0}$ is the orientation
distribution function of the parent nematic phase. This assumption is already
suspect on purely physical grounds, since we expect enhancement of the nematic
order inside the smectic layer due to the increased local density there. This
suspicion is confirmed if we apply the general form of the first bifurcation
equation \rf{eq:bif1} to this problem. We find that the bifurcating solution
has the initial form
\be
\psi_{1}({\bf r},{\bf n}) = \phi_{{\bf q}}({\bf n}) \cos 
          {\bf q \cdot r}
\ee
where the function $\phi_{{\bf q}}$, which describes the lowest order response
of the orientational distribution to the smectic density wave, is a solution of
the following equation
\be
\phi_{{\bf q}}({\bf n}) = -n\psi_{0}({\bf n}) \int \!d{\bf n}' 
          \hat{c}^{(2)}_{0}({\bf q},{\bf n},{\bf n}')\phi_{{\bf q}}({\bf n}')
\ee
which contains the fourier transform $\hat{c}^{(2)}_{0}$ of the direct pair
correlation in the nematic phase. Although linear, this is a highly non-trivial
equation, mainly because its kernel is, even in the Onsager approximation, a
complicated function of its arguments. Solving it would, among other things,
give a first principles demonstration that the smectic density wave is indeed
parallel to the nematic director.

The second interesting development is the evidence for the existence of a so
called cubatic phase in a system of hard cut-spheres by Frenkel and co-workers
\cite{fren:stat,veer:cuba}. This is a phase, which possibly has long range
cubic orientational order without, however, long-range positional order (it is
of course highly structured locally). Not only is the type of order new but it
also surprising that uniaxially symmetric bodies like the cut-spheres can form
homogeneous phases of lower symmetry, in this case that of the cubic group
$O_{h}$. Since the evidence suggest that this phase develops spontaneously from
the isotropic phase without any intermediate nematic, it would seem feasible to
understand its creation using bifurcation analysis of a suitable functional. As
a preparatory exercise we are currently studying a model system that is
guaranteed to show a cubatic phase, albeit through a different mechanism. This
model is that of the so called Onsager crosses, introduced by Frenkel
\cite{fren:stat} to study the possibilities of the formation of liquid crystals
with exotic symmetries.  The model consist of particles composed of three
mutually orthogonal infinitely thin hard rods that are connected to each other
in their centers of mass so as to form a rigid cross.

\section{Conclusions}

At the end of this short guide to the application of bifurcation analysis to
the study of liquid crystal phase transitions, it seems fitting to put the
technique once more in perspective. First of all it is useful to bear in mind
that it is indeed a {\em tool\/} and {\em not} a theory. In the context
discussed  here, formulating a theory is equivalent to specifying a free energy
functional. There is no general recipe for this process of theory formation
although we are guided by criteria like simplicity and unbiasedness (you should
take care not to put in by hand what you want to get out !). The Onsager
approximation for hard non-spherical particles scores well on these points,
which, apart from the fact that it also yields interesting results, accounts
for its ongoing and well-deserved use. Bifurcation analysis, on the other hand,
is just a technique for obtaining some of the consequences of a given theory by
analyzing the non-linear equations that describe the predicted equilibrium
phases. It is, however, a rather powerful technique, and I hope to have given
the reader some impression of this in these lectures. More specifically it
focusses exclusively on the most interesting aspect of any theory viz. its
predicted phase transitions and their properties. If a theory can be compared
to an oyster, then bifurcation analysis is one of those smart little implements
that break it open in order to get at the pearl, being the phase transition.
Given the continued activity in the field  and the ever increasing knowledge
obtained by computer simulations on well defined model systems, I feel
confident that bifurcation analysis is just at the beginning of its ``product
life-cycle'' in liquid crystal research.

\section*{Acknowledgments}
I would like to thank the director of this Advanced Research Workshop for
inviting me to lecture and the NATO for covering the expenses. I would also
like to express the hope that this last-mentioned institution will in the near
future be able to occupy itself exclusively with cultural and scientific
cooperation by implementing its own version of the ``swords to ploughshares''
program.  Finally, I am indebted to the Institute for Materials Science of the
National Center for Scientific Research ``Demokritos'' for their financial
support during the preparation of the manuscript.

\end{document}